\documentclass{Interspeech2024}

\interspeechcameraready

\usepackage{microtype}
\usepackage[noabbrev]{cleveref}
\usepackage{fancyvrb}
\usepackage{cite}

\title{Spoken-Term Discovery using Discrete Speech Units}

\name[affiliation={1}]{Benjamin}{van Niekerk}
\name[affiliation={2}]{Julian}{Zaïdi}
\name[affiliation={2}]{Marc-André}{Carbonneau}
\name[affiliation={1}]{Herman}{Kamper}

\address{
  $^1$E\&E Engineering, Stellenbosch University, South Africa\\
  $^2$Ubisoft La Forge, Montreal, Canada
}
\email{benjamin.l.van.niekerk@gmail.com, julian.zaidi@ubisoft.com, marc-andre.carbonneau@ubisoft.com, kamperh@gmail.com}


\begin{document}

\maketitle
 
\begin{abstract}
Discovering a lexicon from unlabeled audio is a longstanding challenge for zero-resource speech processing. 
One approach is to search for frequently occurring patterns in speech. 
We revisit this idea with DUSTED: Discrete Unit Spoken-TErm Discovery\footnote{\scriptsize Code available at \url{https://github.com/bshall/dusted}}.
Leveraging self-supervised models, we encode input audio into sequences of discrete units.
Next, we find repeated patterns by searching for similar unit sub-sequences, inspired by alignment algorithms from bioinformatics.
Since discretization discards speaker information, DUSTED finds better matches across speakers, improving the coverage and consistency of the discovered patterns. 
We demonstrate these improvements on the ZeroSpeech Challenge, achieving state-of-the-art results on the spoken-term discovery track.
Finally, we analyze the duration distribution of the patterns, showing that our method finds longer word- or phrase-like terms.
\end{abstract}

\section{Introduction}

Spoken-term discovery aims to find recurring speech segments representing words or
short phrases.
The main difficulty is the enormous variability of spoken language. 
Words are seldom said the same way due to differences in speaking rate, intonation, pronunciation, context, and speaker identity. 
Another challenge is segmentation---delineating continuous speech into separate words~\cite{saffran_statistical_1996}. 
Unlike the spaces between written words, speech rarely has easily identifiable boundaries.
Despite this complexity, children learn to recognize a few words even before their first birthday~\cite{bergelson_at_2012}.  
Their vocabulary expands rapidly over the next years, growing to about a thousand words by age three~\cite[p.282]{shipley_assessment_2023}.

Recently, the ZeroSpeech Challenge~\cite{dunbar_zero_2020} has driven progress on this problem.
The goal is to build systems that generalize across languages without requiring textual annotations or labels.
Such systems could facilitate low-resource speech technology~\cite{kamper_segmental_2017} or serve as cognitive models of language acquisition~\cite{dupoux_cognitive_2018}.

Although various methods have been developed to tackle spoken-term discovery~\cite{rasanen_unsupervised_2015, kamper_embedded_2017, bhati_self-expressing_2020}, many submissions to the Zero-Speech Challenge rely on dynamic time-warping (DTW)~\cite{park_unsupervised_2008, jansen_efficient_2011, lyzinski_evaluation_2015, rasanen_unsupervised_2020}.
These methods trace back to the Segmental-DTW algorithm~\cite{park_unsupervised_2008}.
The basic idea is to search for similar speech patterns by aligning pairs of utterances using DTW.
Intuitively, shared words between the utterances will sound similar, leading to low-distortion regions in the alignment.

However, DTW-based methods have several drawbacks. 
Older methods search for recurring patterns by exhaustively aligning every pair of utterances in a dataset. 
But, increasing dataset sizes have made this impractical. 
Instead, modern methods rely on heuristics such as pre-filtering and windowing to manage
computational costs~\cite{jansen_efficient_2011}.
Additionally, alignments are typically computed on spectral features that contain speaker-specific information. 
As a result, it is difficult to find matching patterns across speakers. 
This can cause DTW-based methods to miss infrequently repeated words.
Finally, it is challenging to set hyperparameters that perform consistently across different datasets and languages~\cite{rasanen_unsupervised_2020}.

To address these limitations, we revisit the idea of pattern matching using discrete speech representations.
Leveraging recent self-supervised speech models, we encode input audio into sequences of discrete units~\cite{hsu_hubert_2021, van_niekerk_comparison_2022}.
Next, we find matching segments across pairs of utterances by searching for common sub-sequences of units.
Since discrete units mainly capture phonetic information, the idea is to find matches based on content rather than speaker-specific details.

We evaluate our method on the spoken-term discovery track of the ZeroSpeech
Challenge. 
Next, we investigate the effect of pre-training language and clustering
strategies. 
Finally, we analyze the speaker composition and duration distribution of
the discovered patterns.

Our main contributions are:
\begin{enumerate}
    \item We propose DUSTED: \textbf{D}iscrete \textbf{U}nit \textbf{S}poken-\textbf{TE}rm \textbf{D}iscovery.
    Our approach significantly increases the number of discovered pairs, particularly across speakers (\Cref{sec:speaker-analysis}).
    \item We investigate the trade-off between the quality and quantity of discovered pairs (\Cref{sec:comparison-existing-systems}). 
    By adjusting a single threshold, we can prioritize coverage or phonetic similarity.  Additionally, we show that similar threshold settings perform consistently across languages, giving state-of-the-art results on the ZeroSpeech 2017 Challenge.
    \item We quantify native language caused by the discrete units by comparing pattern matching on one language using units learned on another~(\Cref{sec:cross-lingual-comparison}).
    In contrast to previous work~\cite{millet2022}, we find that the units are not language-independent. Instead, targeting a specific language improves performance.
\end{enumerate}



\section{Method}

DUSTED consists of two parts.
First, the content encoder extracts discrete representations of speech.
Next, the pattern matcher finds candidate words by searching for similar speech segments across pairs of utterances.
 
\subsection{Content Encoder}
\label{sec:content-encoder}

\begin{figure}[!t]
    \centering
    \includegraphics[width=\columnwidth]{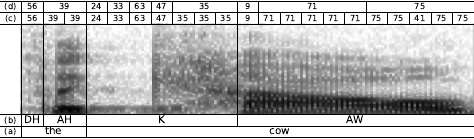}
    \caption{
    \textbf{Content Encoder}. An example segmentation of the phrase `the cow'.
    a) Ground truth word boundaries.
    b) Aligned phonetic transcription.
    c) Discrete speech units extracted by clustering features from an intermediate layer of HuBERT.
    d) A grouping of the units into longer segments using the method described in \Cref{sec:content-encoder}.
    }
    \label{fig:segmentation}
    \vspace{-4pt}
\end{figure}

The content encoder extracts discrete speech representations that discard speaker
information~\cite{vanniekerk2021analyzing}.
Reducing variation across speakers is crucial for matching patterns based on content.
For the same reason, discrete units are useful for voice conversion~\cite{polyak_speech_2021, van_niekerk_comparison_2022} and speech-to-speech translation~\cite{lee_textless_2022}.
Here, we discretize input speech by clustering features from an intermediate layer of HuBERT~\cite{hsu_hubert_2021}.
Formally, given a sequence of features $\langle \mathbf{z}_1, \ldots, \mathbf{z}_T \rangle$, we replace each frame with the index of the nearest cluster centroid.
\Cref{fig:segmentation}(c) illustrates this step.

Often, neighboring frames belong to the same cluster.
Nevertheless, some acoustically similar frames are mapped to different units.
For instance, the end of the vowel $/\text{AW}/$ in \Cref{fig:segmentation}(b) is split between clusters 75 and 41.
So, to group the frames into longer segments we apply the dynamic programming method from \cite{kamper_towards_2021}.
Specifically, we partition the frames into a sequence of contiguous segments $\langle g_1, \ldots, g_N \rangle$, where each segment $g_n = (a_n, b_n, i_n)$ is defined by a start step $a_n$, an end step $b_n$, and a representative cluster index $i_n$.
We determine the segmentation by minimizing the total distance between the features and their assigned cluster centroids:
\[
\mathcal{E}(\mathbf{z}_{1:T}, g_{1:N}) = \sum_{g_n \in g_{1:N}} \sum_{t=a_n}^{b_n} \lVert \mathbf{z}_t - \mathbf{e}_{i_n} \rVert - \gamma (b_n - a_n),
\]
where $\mathbf{e}_i$ is the $i$th centroid.
The last term in the summation encourages longer segments, with $\gamma$ controlling its weight.
\Cref{fig:segmentation}(d) shows an example segmentation where the units in row (c) are combined into longer groups.
Ultimately, the content encoder represents an utterance as the sequence of cluster indexes given by the segmentation.

\subsection{Pattern Matcher}
\label{sec:pattern-matcher}

\begin{figure}
    \centering
    \includegraphics[width=\columnwidth]{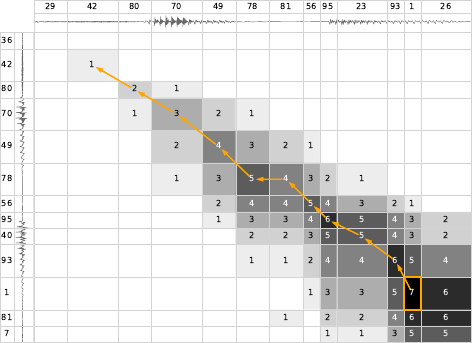}
    \caption{
    \textbf{Pattern Matcher}. The scoring matrix and alignment path for two instances of the word `something'.
    The first row and column show discrete representations of the words (obtained from the content encoder).
    The highest score (highlighted in orange) represents the similarity of the aligned sub-sequences.
    The orange arrows visualize the traceback path.
    }
    \label{fig:pattern-matching}
\end{figure}

After translating input speech into discrete units, the pattern matcher searches for similar fragments across pairs of utterances.
The intuition is that matching fragments should represent common words or phrases.
Specifically, we find the most similar sub-sequence given discrete representations for two utterances $\langle x_1, \ldots, x_N \rangle$ and $\langle y_1, \ldots, y_M \rangle$. 
We identify similar sub-sequences using the Smith-Waterman algorithm \cite{smith_identification_1981}, originally designed for nucleic acid or protein sequence alignment.
The algorithm accounts for variability in the sequences by allowing for insertions, deletions, and substitutions.
\Cref{fig:pattern-matching} shows an example alignment using the algorithm. 
The orange path represents the most similar sub-sequence between the two utterances, which includes a gap and a substitution (in bold):
\begin{center}
\begin{BVerbatim}[commandchars=\\\{\}]
 Top: 42 80 70 49 78 \textbf{81} 56 95 \textbf{23} 93 1
Left: 42 80 70 49 78 -- 56 95 \textbf{40} 93 1
\end{BVerbatim}
\end{center}

We score the similarity of sub-sequences based on how many units they have in
common.
We apply the pattern matcher to each pair of utterances in a dataset and record matches scoring above a similarity threshold $\tau$. 
The threshold controls the trade-off between the quantity and quality of the discovered patterns (see the experiments in \Cref{sec:comparison-existing-systems}).

Next, we describe the four steps of the algorithm:
\begin{enumerate}
    \item \textbf{Determine a scoring scheme.}
    First, we define a substitution function $\text{sim}(x, y)$ that returns a score for matching units $x$ and $y$.
    This score is positive if $x$ and $y$ are similar and negative if dissimilar.
    In this paper, we only consider exact matches:
    \[
    \text{sim}(x, y) = \begin{cases}
        +1, \, \text{ if } x=y, \\
        -1, \, \text{ if } x \neq y.
    \end{cases}
    \]
    However, this formulation allows more flexible measures of similarity. 
    For example, we could specify different scores for matching units representing sonorants, obstruents, or silences~\cite{van_niekerk_rhythm_2023}.
    We also define a gap penalty $W$ for including an insertion or deletion in the alignment.
    We set $W=1$ for all experiments. 
    
    \item \textbf{Fill the scoring matrix.}
    Next, we set up a scoring matrix $H$ of size $(N+1)\times(M+1)$.
    The cell $H_{i, j}$ represents the maximum similarity between two sub-sequences ending in $x_i$ and $y_j$.
    We initialize the first row and column of $H$ to zeros and iteratively fill the matrix from left to right and top to bottom using the recurrence:
    \[
    H_{i, j} = \max
    \begin{cases}
        H_{i-1, j-1} + \text{sim}(x_i, y_j), \\
        H_{i-1, j} - W, \\
        H_{i, j-1} - W, \\
        0
    \end{cases} 
    \]
    The first line is the score for aligning $x_i$ with $y_j$.
    The second and third lines account for an insertion or deletion.
    Finally, the zero represents no similarity between the sub-sequences.
    \Cref{fig:pattern-matching} shows the scoring matrix for the sequences along the top and left.

    \item \textbf{Traceback to find the most similar sub-sequence.}
    The traceback starts at the highest-scoring element in $H$ above the similarity threshold $\tau$ (highlighted in orange in \Cref{fig:pattern-matching}).
    If two or more elements are tied for the maximum, we select the one with the lowest index sum $i + j$ (towards the top-left corner in \Cref{fig:pattern-matching}).
    From this starting point, we recursively visit the neighboring element leading to the maximum score. 
    We stop the procedure when we encounter a zero. 
    The orange arrows illustrate the traceback path in \Cref{fig:pattern-matching}.

    \item \textbf{Iteratively identify all matching sub-sequences.} 
    The scoring matrix may include multiple matches above the similarity threshold $\tau$.
    We use the rescoring method from~\cite{waterman_new_1987} to find the next highest-scoring alignment. 
    To avoid overlapping matches, we set all cells along the previous traceback path to zeros and recompute the scoring matrix.
    Just part of $H$ needs to be updated since only elements below and to the right of the path are affected.
    We repeat the traceback and rescoring steps (3 and 4) until no matches above the threshold remain.
\end{enumerate}

\begin{figure}[!t]
    \centering
\includegraphics[width=\columnwidth]{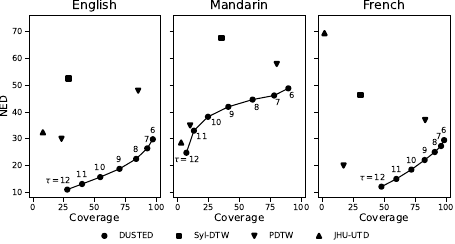}
    \caption{\textbf{Comparison with the Baselines}. Coverage versus NED for DUSTED (at different similarity thresholds $\tau$) and three state-of-the-art systems built on dynamic time-warping.}
    \label{fig:ned-cov}
\end{figure}

\section{Experimental Setup}

We conduct four experiments to evaluate DUSTED.
First we compare DUSTED to three state-of-the-art systems built on dynamic time-warping: PDTW~\cite{rasanen_unsupervised_2020}, Syl-DTW~\cite{rasanen_zs2017_2017}, and JHU-UTD~\cite{jansen_efficient_2011}.
Next, we examine the effect of pre-training language.
Specifically, we investigate pattern matching on one language using discrete units learned on another.
Then, we explore the importance of speaker-invariance by analyzing the impact of discrete units on cross-speaker matches.
Finally, we examine the duration distribution of the discovered patterns, showing that DUSTED finds longer word- or phrase-like terms.

We evaluate DUSTED on the spoken-term discovery track of the ZeroSpeech Challenge~\cite{dunbar_zero_2020}. 
The challenge covers five languages: English, Mandarin, French, German, and Wolof.
We limit our experiments to languages with publicly available HuBERT models (English\footnote{\tiny \url{https://huggingface.co/facebook/hubert-base-ls960}}, Mandarin\footnote{\tiny \url{https://huggingface.co/TencentGameMate/chinese-hubert-base}}, and French\footnote{\tiny \url{https://huggingface.co/voidful/mhubert-base}}).
We were unable to find a language-specific model for French. 
So we use a multilingual model trained on French, English, and Spanish~\cite{lee_textless_2022}.

\subsection{Implementation Details}
\label{sec:implementation-details}

We split the evaluation datasets into short audio clips using the voice activity detection markers provided by the challenge.
Then, we extract features for each language using the corresponding HuBERT model.
Following previous work~\cite{van_niekerk_comparison_2022}, we take activations from the 7th transformer layer because they perform well for phone discrimination~\cite{hsu_hubert_2021, nguyen2020zero}.
We cluster the features using $k$-means with 100 clusters.
Next, we apply the method described in \Cref{sec:content-encoder} to segment the features into phone-like units (setting the duration weight to $\gamma=0.2$, following~\cite{van_niekerk_rhythm_2023}).
Finally, we find matching patterns between each pair of utterances in a given language dataset using
the method from \Cref{sec:pattern-matcher}.
We filter out short matches that are unlikely to contain complete words.
Specifically, we ignore matches below \SI{200}{\milli\second} given that the average duration of a consonant-vowel syllable is \SI{156}{\milli\second}~\cite{kuwabara1996}.
We report results at thresholds $\tau$ from 6 to 12.

\subsection{Evaluation Metrics}

We evaluate spoken-term discovery using the matching metrics provided by the ZeroSpeech Challenge.
The first metric is coverage: the proportion of the corpus covered by the patterns (higher is better).
The second is normalized edit distance (NED), which measures the phonetic similarity between discovered pairs.
Computing NED requires time-aligned transcriptions for each discovered pattern. 
A phone is included in a transcription if it overlaps with the pattern by more than \SI{30}{\milli\second} or 50\% of its duration.
Then, we calculate the normalized Levenshtein distance between the transcriptions of each discovered pair.
Finally, NED reports the average distance over all pairs (lower is better).

\section{Results}

\begin{figure}
    \centering
\includegraphics[width=\columnwidth]{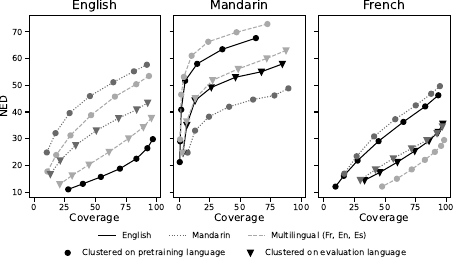}
    \caption{\textbf{Effect of Language Pre-training and Clustering}.
    We compare pattern matching on one language using discrete units learned on another.
    We report coverage versus NED at different similarity thresholds $\tau$.
    }
    \label{fig:ned-cov-crosslingual}
\end{figure}

\subsection{Comparison to State-of-the-Art Systems}
\label{sec:comparison-existing-systems}

This section compares DUSTED to existing methods based on dynamic time-warping.
Typically, spoken-term discovery balances NED against coverage. 
DUSTED controls this trade-off through the similarity threshold $\tau$.
Increasing the threshold encourages longer, more similar matches.
However, being more restrictive leads to fewer pairs and lower coverage.
We further investigate the threshold's effect on the duration of the discovered patterns in \Cref{sec:duration-analysis}.

\Cref{fig:ned-cov} reports the performance of DUSTED alongside three state-of-the-art methods.
The ideal system would be in the bottom-right corner (low NED and high coverage).
Regardless of the threshold, DUSTED outperforms other methods operating at similar trade-off points.
At comparable coverage, we improve NED over PDTW by $13.5$ points on average.
Additionally, the threshold's effect is relatively consistent across languages, allowing us to reliably prioritize NED or coverage.

A drawback of DUSTED is the amount of data required to train the content encoder on new languages~\cite{hsu_hubert_2021}.
While DTW-based methods use spectral features, we rely on self-supervised models trained on large datasets. 
One method to address this limitation is transfer learning from a model trained on well-resourced languages.
We analyze the effect of language transfer in the next section.

\subsection{Effect of Language Pre-training and Clustering}
\label{sec:cross-lingual-comparison}

We investigate the native language effect of the content encoder in two scenarios: 
\begin{enumerate}
    \item The training language of the content encoder and $k$-means clustering differs from the evaluation language. 
    For example, we could use an English HuBERT clustered on English data to encode French speech.
    
    \item We cluster on the evaluation language, but the content encoder is trained on a different language.
    Here, we would use an English HuBERT but cluster on French data. 
\end{enumerate}
Scenario 1 represents the largest mismatch between the content encoder and evaluation language. 
We test all combinations of training and evaluation languages using the hyperparameters described in \Cref{sec:implementation-details}.

\Cref{fig:ned-cov-crosslingual} presents our findings.
Overall, matching the training and evaluation languages leads to the best performance.
Compared to the mismatched content encoders (other lines with circle markers), the performance discrepancy suggest that HuBERT learns language-specific representations, contradicting previous work~\cite{millet2022}.
However, clustering on the evaluation language (triangle markers) improves performance despite a mismatched content encoder, showing we can mitigate some language misalignment.

The results for the multilingual content encoder are particularly interesting.
Although the pre-training languages include English, the multilingual encoder performs worse than the English-specific model.
Additionally, when evaluating on Mandarin, multilingual training gives no advantage over training solely on English.
While \cite{conneau_unsupervised_2021} argues that multilingual training results in transferable representations~\cite{conneau_unsupervised_2021}, our experiments do not show this advantage.
To summarize, matching the pre-training language to the evaluation language gives the best results.

\subsection{Analysis of Speaker Invariance}
\label{sec:speaker-analysis}

\begin{figure}
    \centering
\includegraphics[width=\columnwidth]{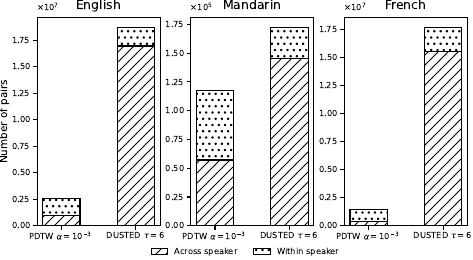}
    \caption{A comparison of the number of within- and across-speaker pairs discovered by PDTW~\cite{rasanen_unsupervised_2020} and DUSTED.}
    \label{fig:pair-count}
\end{figure}

This section analyzes the speaker composition of the discovered patterns.
\Cref{fig:pair-count} compares the number of pairs found by DUSTED and PDTW, divided into across-speaker and within-speaker matches.
In line with the coverage results from \Cref{sec:comparison-existing-systems}, DUSTED discovers more patterns in each language.
Importantly, DUSTED predominantly finds pairs from different speakers: over $80\%$ of the matches are cross-speaker, compared to less than $50\%$ for PDTW.
These findings demonstrate that the discrete speech units effectively discard speaker information.
As a result, the pattern matcher can discover terms based on content rather than speaker-specific details.
This is essential for spoken-term discovery since many words and phrases will not be repeated by the same speaker.
In contrast, PDTW relies on spectral features that contain speaker information, limiting the number of cross-speaker matches.

\subsection{Duration of Discovered Fragments}
\label{sec:duration-analysis}

\begin{figure}
    \centering
\includegraphics[width=\columnwidth]{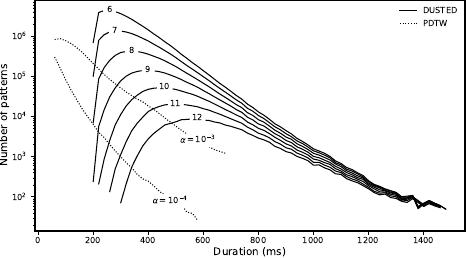}
    \caption{The duration distribution of discovered patterns on English for DUSTED (at different similarity thresholds $\tau$) and PDTW~\cite{rasanen_unsupervised_2020} (at different significance thresholds $\alpha$).}
    \label{fig:duration-distribution}
\end{figure}

Finally, we examine the durations of the discovered patterns.
Ideally, the patterns should capture words or short phrases spanning hundreds of milliseconds to over a second.
\Cref{fig:duration-distribution} shows duration distributions for DUSTED and PDTW at different thresholds.
As discussed in \cref{sec:comparison-existing-systems}, raising the threshold $\tau$ encourages longer matches with higher similarity, reflected in a larger average duration of the patterns.
However, more restrictive thresholds reduce the number of matches, lowering overall coverage.

\Cref{fig:duration-distribution} shows that DUSTED discovers longer fragments than PDTW.
To reduce computational costs, PDTW imposes a maximum window size on alignments, limiting the length of the discovered patterns to \SI{700}{\milli\second}.
Consequently, PDTW discovers shorter fragments concentrated around 100 ms---roughly the duration of a syllable~\cite{kuwabara1996}.
On the other hand, DUSTED does not set an upper limit and discovers patterns ranging from 200 to \SI{1400}{\milli\second}.

\section{Conclusion}

This paper introduced DUSTED, a new spoken-term discovery method combining pattern matching with discrete speech units.
Since discrete units discard speaker information, DUSTED finds matches based on phonetic content rather than speaker details.
This results in significantly better coverage, particularly across speakers. 
Our experiments showed that DUSTED outperforms existing systems on the ZeroSpeech Challenge, improving the quality and quantity of the discovered terms.
We also evaluated the impact of pre-training language on the discrete speech units. 
Our findings indicate that self-supervised representations are not language-independent, and that language-specific models can improve spoken-term discovery.

\section{Acknowledgements}

We thank Okko Räsänen for kindly providing us with the official submission files for PDTW.

\bibliographystyle{IEEEtran}
\bibliography{custom}

\end{document}